\documentclass[twocolumn, amsmath, amssymb, superscriptaddress, prb]{revtex4-1}
\usepackage{mathtools}
\mathtoolsset{showonlyrefs=true}
\usepackage{graphicx}
\usepackage{bm}
\usepackage{here}
\usepackage{url}
\usepackage{ulem}
\usepackage{xcolor}
\begin{document}
\title{Effective Ruderman-Kittel-Kasuya-Yosida-like Interaction in Diluted Double-exchange Model: Self-learning Monte Carlo Approach}
\author{Hidehiko Kohshiro}
\affiliation{Institute for Solid State Physics, The University of Tokyo, Kashiwa, Chiba, Japan}
\affiliation{CCSE, Japan Atomic Energy Agency, 178-4-4, Wakashiba, Kashiwa, Chiba, 277-0871, Japan}
\author{Yuki Nagai}
\affiliation{CCSE, Japan Atomic Energy Agency, 178-4-4, Wakashiba, Kashiwa, Chiba, 277-0871, Japan}
\affiliation{Mathematical Science Team, RIKEN Center for Advanced Intelligence Project (AIP), 1-4-1 Nihonbashi, Chuo-ku, Tokyo 103-0027, Japan}
\date{\today}

\begin{abstract} 
    We study the site-diluted double-exchange (DE) model and its effective Ruderman-Kittel-Kasuya-Yosida-like interactions, where localized spins are randomly distributed, with the use of the Self-learning Monte Carlo (SLMC) method. The SLMC method is an accelerating technique for Markov chain Monte Carlo simulation using trainable effective models. We apply the SLMC method to the site-diluted DE model to explore the utility of the SLMC method for random systems. We check the acceptance rates and investigate the properties of the effective models in the strong coupling regime. The effective two-body spin-spin interaction in the site-diluted DE model can describe the original DE model with a high acceptance rate, which depends on temperature and spin concentration. These results support a possibility that the SLMC method could obtain independent configurations in systems with a critical slowing down near a critical temperature or in random systems where a freezing problem occurs in lower temperatures.
\end{abstract}
\maketitle

\section{Introduction}
Interactions between itinerant electrons and localized spins have attracted much attention in magnetic materials such as the Manganite~\cite{Zener, AndersonHasegawa, Kubo, Kaplan, ManganitesRMP}.
The double-exchange (DE) model, a fundamental model of the itinerant magnetism, has also attracted much attention from theoretical and experimental points of view since its itinerant nature yields various rich phases~\cite{Nagai, SanjeevPinaki, David, SanjeevJeroen, Minami,AkagiMotome}.

The electron-spin interaction in weak coupling regime in the DE model is approximated by the Ruderman-Kittel-Kasuya-Yosida (RKKY) interaction between localized spins, whose coupling constant oscillates and decays algebraically with the distance between localized spins~\cite{RudermanKittel, Kasuya, Yoshida}. 
In the diluted magnetic alloys (\textit{e.g.}, AuFe and CuMn)~\cite{Mydosh1}, the spin-glass (SG) order, where spins freeze in the spatially random manner, is explained by the existence of the RKKY interaction between randomly localized spins.
The SG state shows novel transport phenomena due to the random spin structure~\cite{TataraKawamura, Niimi}. 
The nature of SG, however, is still under debate even if it is treated as a simplified classical spin model, the so-called Edwards-Anderson model, whose exchange coupling constants are given by a certain probability distribution~\cite{BinderYoung, Mydosh2, KawamuraBook}.

Even treating localized spins as classical ones, the studies beyond the weak coupling regime are not trivial since itinerant electrons play a more important role in the DE model due to its quantum nature.
For example, the existence of the SG order in the site-diluted DE model has not been theoretically confirmed yet.
A standard method to simulate classical spins is the Markov chain Monte Carlo (MCMC) method. 
In the DE model, however, computational complexity is huge due to a fermion determinant for itinerant electrons.
For example, this naive calculation takes $\mathcal{O}(N^4)$ per sweep with the system size $N$ since a full diagonalization is needed to estimate the fermion determinant for given spin configurations for each Monte Carlo (MC) step. 
Many alternative algorithms to simulate electrons coupled to classical fields have been developed~\cite{Alonso, Furukawa, FurukawaMotome, AlvarezSen, AlvarezPhani, Weisse, BarrosKato}.

In MCMC methods for physics, obtaining independent configurations sometimes becomes hard near a critical temperature or at low temperatures. 
A critical slowing down of the autocorrelation near the critical temperature is an infamous problem when the update algorithm for generating configurations is local. 
Global update algorithms such as the Wolff~\cite{Wolff} and the Swendsen-Wang cluster algorithms~\cite{SwendsenWang} are usually adopted for reducing the autocorrelation time 
if these algorithms are available. 

The freezing problem is another infamous problem caused by the multi-valley structure of the energy landscape manifested at low temperatures, and the states are trapped in the local minima. 
The freezing problem appears in random systems such as the SG systems and makes difficult its MCMC simulations.
In principle, one can reach the true minima in finite-temperature systems after many MC steps. 

Recently, an efficient MCMC algorithm combined with a machine-learning technique, the self-learning Monte Carlo (SLMC) method, was proposed~\cite{SLMCIsing}. 
The SLMC method constructs effective models to generate the Boltzmann weight by means of machine learning from gathered original Markov chains.
Once an effective model is prepared, we can perform MCMC simulations of the original models using the effective models trained. 
It was shown that SLMC reduces the number of the MC steps of the original model even near the critical slowing down~\cite{SLMCIsing, SLMCDE}.
Using an effective model where cluster algorithms are available such as the Ising model, the SLMC method accelerates a simulation of a model where cluster algorithms are unavailable~\cite{SLMCIsing}.
As shown in Fig.~\ref{fig:alg}, with the use of the SLMC, we can bypass heavy numerical computations coming from an estimation of the original model. 
This property is also useful for solving the freezing problem. 

It is not trivial whether the SLMC method is efficient for random systems. 
If one uses a very complicated effective model with deep neural networks, the original model might be reproduced by the effective model. 
However, in terms of the computational complexity of actual MCMC simulations, a simple effective model is needed to bypass heavy numerical computations. 
In the week coupling regime, there is a simple effective model known as the RKKY interaction, which was also used for the randomly distributed spin systems~\cite{WalstedtWalker,Zhang}. 

In this paper, we show that there is a simple effective model similar to the RKKY interaction even in the strong coupling regime. 
To prove it, We employed the SLMC method with the cumulative update for fermions coupled to the classical field~\cite{SLMCDE}. We first checked success or failure through the acceptance rate. Next, we investigated the properties of the effective models trained. The main issues is the dependence of temperature and distance between localized spins of the effective models trained. The temperature dependence was not investigated in the site-diluted DE model and also in the regular DE model previously.
 
This paper is organized as follows. First, we introduce the site-diluted DE model. Next, we explain how to perform the SLMC simulation and show the obtained results for the site-diluted DE model. Finally, we summarize and give some discussions.

\section{Model and Method}
\begin{figure}
 \includegraphics[width=\columnwidth]{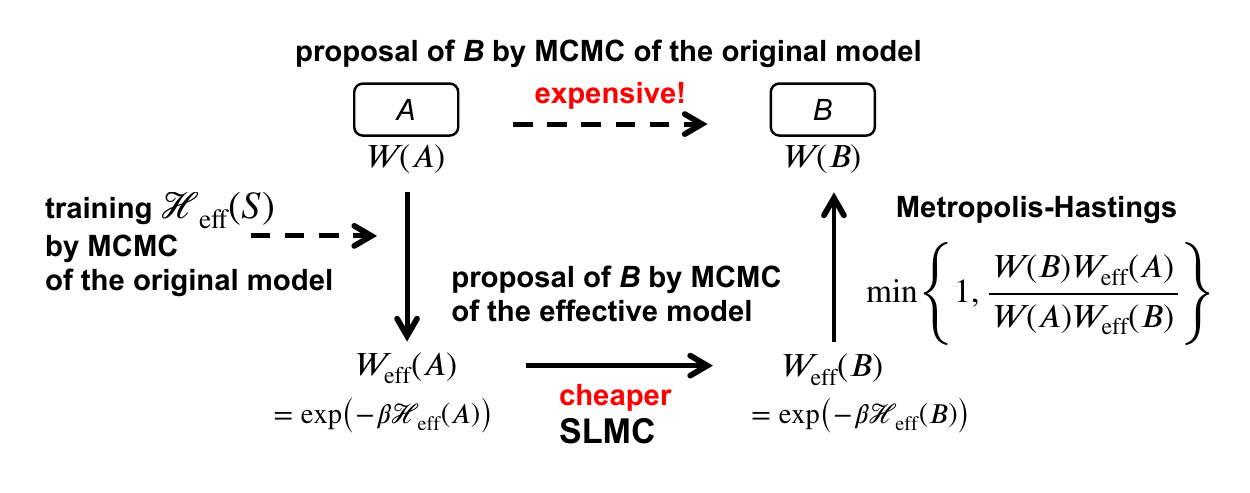}
 \caption{\label{fig:alg}(Color online) A brief schematic summary of the SLMC procedure. The computational cost of the SLMC method is cheaper than that of the naive Monte Carlo method.
 }
 \end{figure}
 
 In this paper, we focus on the site-diluted DE model on the three-dimensional cubic lattice to examine the efficiency of the SLMC method. 
 The site-diluted DE model (the site-diluted ferromagnetic Kondo model) is defined as follows:
 \begin{align}
 \hat{\mathcal{H}}=-t\sum_{\sigma, \langle i,j \rangle}\left(\hat{c}^\dagger_{i\sigma}\hat{c}^{}_{j\sigma}+\mathrm{h.c.}\right)
 -\frac{J}{2}\sum_ip_i\bm{S}_i\cdot\hat{\bm{\sigma}}_i-\mu \sum_{\sigma, i}\hat{c}^\dagger_{i\sigma}\hat{c}^{}_{i\sigma},
 \end{align}
 where $\bm{S}_i$ and $\hat{\bm{\sigma}}_i$ are the classical ($S=\infty$) Heisenberg spins and the Pauli matrices respectively. 
 $\bm{S}_i$ is normalized, $|\bm{S}_i|^2=1$.
 $t>0$ and $J>0$ are the hopping and coupling constant. $p_i$ is a binary random number (0 or 1) which represents whether the site $i$ is occupied ($p_i = 1$) or not ($p_i = 0$) with a localized spin. When $p_i$ is always 1 for an arbitrary $i$, The model corresponds to the regular DE model. It is known the regular DE model holds the ferromagnetic order for enough large $J/t$~\cite{AlvarezSen, MotomeFurukawa}. For simplicity, we utilize a semiclassical approximation where localized spins are classical ones, and energy of the electrons changes instantaneously with the change of the localized spins \textit{i.e.,} the localized spins are ``\textit{time-independent}''.
 
 The partition function of the model $\mathcal{Z}$ is given by
 \begin{align}
 \mathcal{Z}&=\sum_{\mathcal{S}}W(\mathcal{S}),
 \end{align}
 where $W(\mathcal{S})\equiv \prod_n \left(1+\mathrm{e}^{\beta\left(\mu-E_n(\mathcal{S})\right)}\right)$ is the Boltzmann weight with a spin configuration $\mathcal{S}$. 
 Here, $\{E_n(\mathcal{S})\}$ is the eigen spectrum.
 
 We show a summary of the procedure of the SLMC simulation in Fig.~\ref{fig:alg}. The SLMC method is based on two procedures: ``learn'' and ``earn''. ``Learn'' is the training procedure of the effective models. We generate $W(\mathcal{S})$ of the original models as the learning data by performing MCMC simulation of the original model. We optimize the effective model by using the generated learning data so that it generates $W_{\mathrm{eff}}(\mathcal{S})$ closer to original $W(\mathcal{S})$. 
We adopt a site-diluted two-body classical spin interaction up to the sixth neighbor, as in the previous study for the regular DE model~\cite{SLMCDE}.
The effective Hamiltonian is expressed as
\begin{align}
 \mathcal{H}_\mathrm{eff}&=E_0-\sum_{\langle i, j\rangle_n} J^\mathrm{eff}_np_ip_j\bm{S}_i\cdot\bm{S}_j,
\end{align}
where, $\{E_0, J_n^{\mathrm{eff}}\}$ are trainable parameters which we optimize. The difference from the previous study~\cite{SLMCDE} is that we only consider the interactions between randomly arranged spins. In the Appendix~\ref{sec:opt}, we explain details for optimizing the effective model.
 
``Earn'' is the actual simulation of the original model using the effective model optimized. 
The acceptance rate from current state $\mathcal{S}$ to a new state generated by the MCMC of the effective model $\mathcal{S}'$ is given by
\begin{equation}
p(\mathcal{S}\to \mathcal{S}')= \min\left\{1, \frac{W(\mathcal{S}')W_\mathrm{eff}(\mathcal{S})}{W(\mathcal{S})W_\mathrm{eff}(\mathcal{S}')}\right\}, \label{eq:p}
\end{equation}
which is known as the Metropolis-Hastings test.
We note that a proposed state $\mathcal{S}'$ is cumulatively updated~\cite{SLMCDE} and the length of the effective Markov chain is not limited. 
We can update states as long as we need to reduce autocorrelations since the acceptance rate is converged as a function of the length of the effective Markov chain. 

In the effective MCMC, we use the heatbath and the overrelaxation procedure, which is known as an effective update algorithm in classical spin systems.
One effective MC step consists of one heatbath sweep and four overrelaxations.
After 200 effective MC steps,
the Metropolis-Hastings test determined by Eq.~\eqref{eq:p} is done with the proposed spin configuration. 
We use an annealing-like training process to obtain effective models with different temperatures. 
We first gather training data set and train the effective model $\mathcal{H}_\mathrm{eff}$ at a higher temperature (at $T=0.4$ in this paper).
To obtain the effective models in lower temperatures, we use the SLMC with the effective model constructed in the higher temperature and update the model using current data. 
 
We set $t$ as unity and $J=16$ and $\mu=-8$, which possess the ferromagnetic order in the regular DE model. 
The condition $\mu=-8$ corresponds to quarter filling in the regular DE model. The filling decreases with decreasing spin concentration, which is discussed in Appendix~\ref{sec:fill}.
We used the three-dimensional cubic lattice $N=4^3$, and we averaged over five different realizations with the fixed spin concentration and investigated with decreasing spin concentrations.

\section{Results}
\subsection{Acceptance Rates}
\begin{figure}
 \includegraphics[width=\columnwidth]{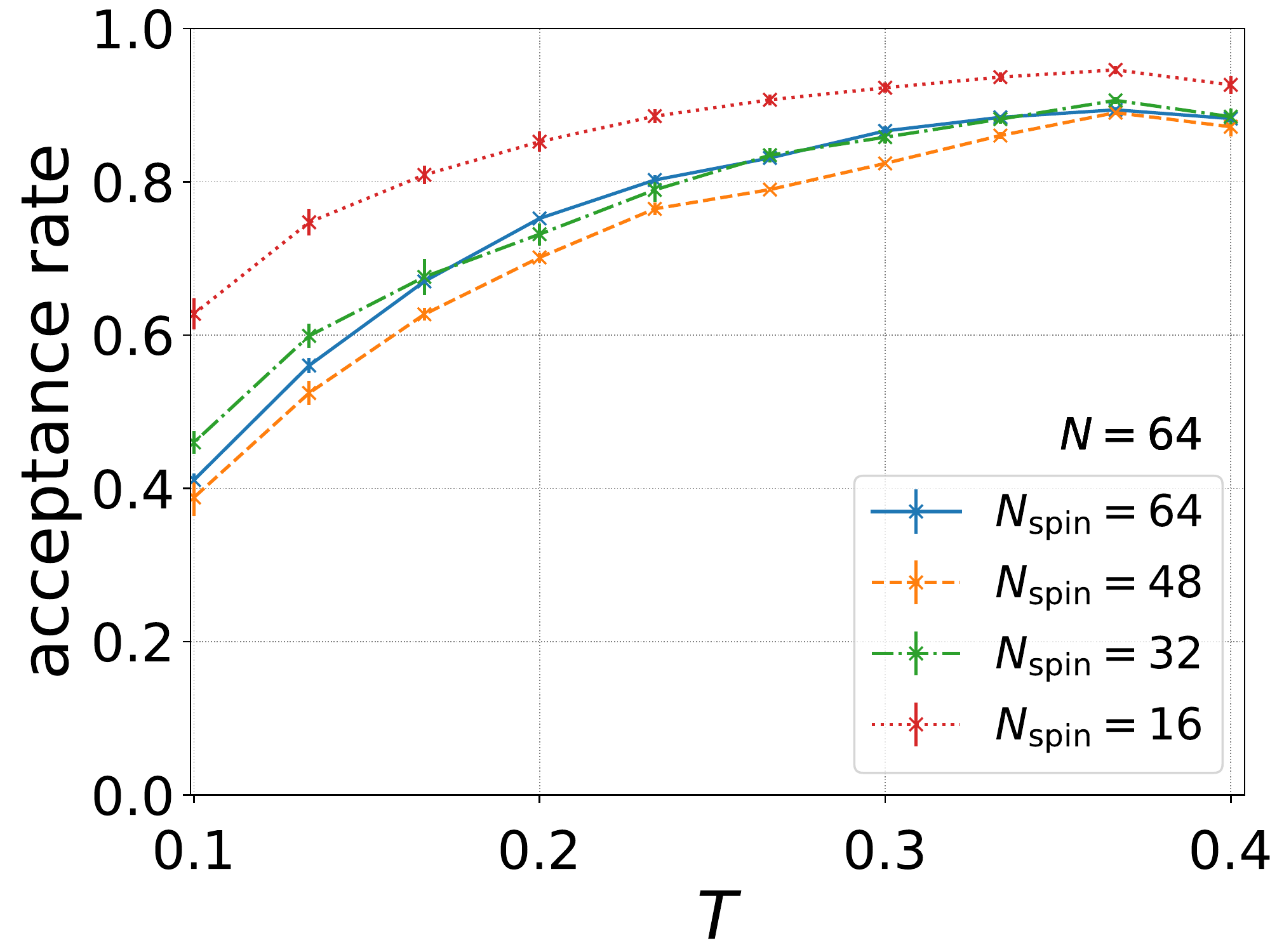}
 \caption{\label{fig:Tacc}(Color online) The acceptance rate versus temperature with different spin concentrations.
 The error bar is defined as the standard error over different localized spin configurations.
 The acceptance rates keep at least 40 \% until $T=0.1$ for all spin concentrations.}
\end{figure}

We investigate the averaged acceptance rate of the SLMC. 
The acceptance rate shows how many proposed spin configurations are accepted during simulations. 
If the learning procedure is failed, the effective model cannot fit the original Boltzmann weight $W(\mathcal{S})$ with a given spin configuration $\mathcal{S}$. 
Such a bad proposal is almost rejected and the acceptance ratio becomes very low, which means a failure of a simulation. 
The averaged acceptance rate is estimated as $p\sim \mathrm{e}^{-\sqrt{\mathrm{MSE}}}$ with the use the mean-squared error (MSE) between the original and effective energy given as~\cite{SLMCDeep}
\begin{align}
\mathrm{MSE} = \frac{1}{N_\mathcal{S}}\sum_\mathcal{S} \left|\log W(\mathcal{S}) - \log W_\mathrm{eff}(\mathcal{S})\right|^2,
\end{align}
where $N_\mathcal{S}$ is the number of spin configurations. 
A high acceptance rate is important for reducing an autocorrelation time~\cite{SLMCNagai}.
We define the autocorrelation time $\tau_\mathrm{A}$ for the observable $A$ in the original MC simulation.
If the number of effective MC steps is larger than $\tau_\mathrm{A}$, the proposed spin configuration is uncorrelated. 
Since the proposed configuration is accepted with the probability $p$, an uncorrelated configuration is obtained after $1/p$ times Metropolis-Hastings tests. 
Therefore, the acceptance rate represents the quality of the effective model and criteria whether the simulation works well or not. 

The averaged acceptance rates are shown in Fig.~\ref{fig:Tacc} in systems with different spin concentrations. 
In the regular model where classical spins are located on all lattice points ($N_\mathrm{spin} = 64$), there is a ferromagnetic transition at $T_c\simeq0.12$~\cite{MotomeFurukawa}. 
We find that there is no drastic change of the acceptance rate around a critical temperature. 
This is because the effective model should not change between different phases. 
The temperature dependence of the acceptance rate is explained by the temperature dependence of the MSE as follows. 
The temperature dependence of the effective action $\log W_\mathrm{eff}(\mathcal{S})$ is described by $\log W_\mathrm{eff}(\mathcal{S},\beta) = -\beta \mathcal{H}_\mathrm{eff}(\beta)$. 
Therefore, the averaged acceptance rate is estimated as 
\begin{align}
p \sim \exp \left( -\sqrt{\beta} \sqrt{\mathrm{MSE}_\mathcal{H}}\right),
\end{align}
where $\mathrm{MSE}_\mathcal{H} \equiv \frac{1}{N_\mathcal{S}}\sum_\mathcal{S}\left|\frac{\log W(\mathcal{S})}{\beta}-\mathcal{H}_\mathrm{eff}(\mathcal{S}, \beta)\right|^2$.
Since the temperature dependence of the acceptance rate shown in Fig.~\ref{fig:Tacc} is well explained by the function $\mathrm{e}^{-a\sqrt{1/T}}$, the mean-squared error of the original and effective Hamiltonian does not depend on the temperature much. 
We note that the effective Hamiltonian itself depends on the temperature as shown in the latter part of this paper.
 
The above analysis can be applied to the cases of other spin concentrations. 
The acceptance rates are kept 40\% below to $T=0.1$. 
This means the SLMC method works well with the site-diluted DE model.
We find that if a system has less localized spins, the acceptance rate becomes high.
This suggests that there is an effective two-body classical spin-spin interaction in the strong coupling regime where $J=16$ in this paper.

\subsection{Spin Concentration and Temperature Dependence of the Effective Models}
\begin{figure}
\includegraphics[width=\columnwidth]{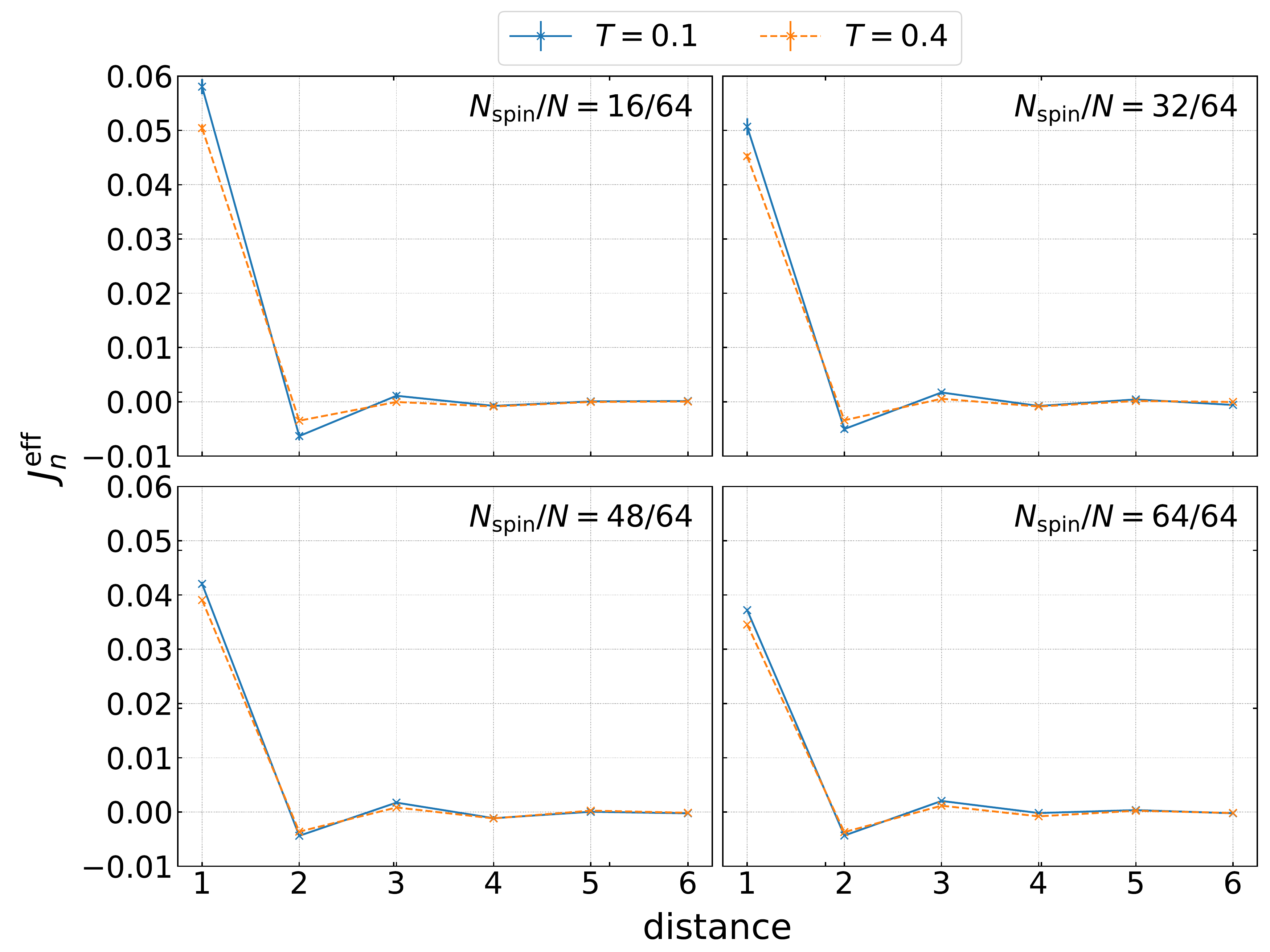}
\caption{\label{fig:rJ}(Color online) The coupling constant $J_n^{\mathrm{eff}}$ of the trained effective models versus distance between localized spins with different spin concentrations at $T=0.1$ and $T=0.4$.
The error bar is defined as the standard error over different localized spin configurations.
 RKKY-like oscillation and decay with distance are observed in all spin concentrations at both $T=0.1$ and $T=0.4$.
 }
\end{figure}

\begin{table*}[t]
\caption{Effective long-range interactions $T =0.4$}
\label{table:1}
\begin{ruledtabular}
\begin{tabular}{cccccccc}
$N_{\mathrm{spin}}$ &
$E_0$ &
$J_1$ &
$J_2$ &
$J_3$ &
$J_4$ &
$J_5$ &
$J_6$
\\64
&-55.253(1)
&0.0691(1)
&-0.00739(9)
&0.00234(5)
&-0.00158(6)
&0.00047(4)
&-0.00033(3)
\\56
&-48.02(1)
&0.0732(2)
&-0.00747(9)
&0.0022(2)
&-0.0020(2)
&0.00047(9)
&-0.00034(5)
\\48
&-40.85(2)
&0.0781(2)
&-0.0072(2)
&0.0017(2)
&-0.0023(1)
&0.00051(9)
&-0.00027(8)
\\40
&-33.77(3)
&0.0837(7)
&-0.0069(2)
&0.00120(6)
&-0.0015(2)
&0.00024(5)
&-0.00003(10)
\\32
&-26.86(3)
&0.091(1)
&-0.0068(2)
&0.0011(1)
&-0.0017(2)
&0.00028(6)
&-0.00006(10)
\\24
&-20.02(2)
&0.095(1)
&-0.0064(4)
&0.0006(2)
&-0.0021(2)
&0.0002(1)
&0.00002(12)
\\16
&-13.36(2)
&0.101(1)
&-0.0070(5)
&-0.00006(42)
&-0.0017(2)
&-0.00006(5)
&0.00009(11)

\end{tabular}
\end{ruledtabular}
\end{table*}
\begin{table*}[t]
\caption{Effective long-range interactions  $T =0.1$}
\label{table:2}
\begin{ruledtabular}
\begin{tabular}{cccccccc}
$N_{\mathrm{spin}}$ &
$E_0$ &
$J_1$ &
$J_2$ &
$J_3$ &
$J_4$ &
$J_5$ &
$J_6$
\\64
&-52.30(1)
&0.0744(7)
&-0.0087(2)
&0.0041(1)
&-0.0004(4)
&0.0007(2)
&-0.0004(2)
\\56
&-45.33(1)
&0.0794(7)
&-0.0099(4)
&0.0043(5)
&-0.0015(2)
&0.0007(2)
&-0.0004(2)
\\48
&-38.38(2)
&0.0841(9)
&-0.0087(5)
&0.0035(5)
&-0.0023(5)
&0.0001(2)
&-0.0005(2)
\\40
&-31.58(6)
&0.090(2)
&-0.0099(9)
&0.004(1)
&-0.0017(8)
&0.0006(1)
&-0.0006(5)
\\32
&-24.97(5)
&0.101(3)
&-0.010(1)
&0.0034(7)
&-0.0015(5)
&0.0009(2)
&-0.0012(2)
\\24
&-18.46(2)
&0.111(2)
&-0.0103(8)
&0.0028(2)
&-0.0025(7)
&0.0012(2)
&-0.0007(2)
\\16
&-12.25(3)
&0.116(3)
&-0.013(2)
&0.002(1)
&-0.0015(9)
&0.0001(4)
&0.0003(4)
\end{tabular}
\end{ruledtabular}
%backup
%0.3(4)$\times 10^{-3}$
\end{table*}

\begin{figure*}
\includegraphics[width=2\columnwidth]{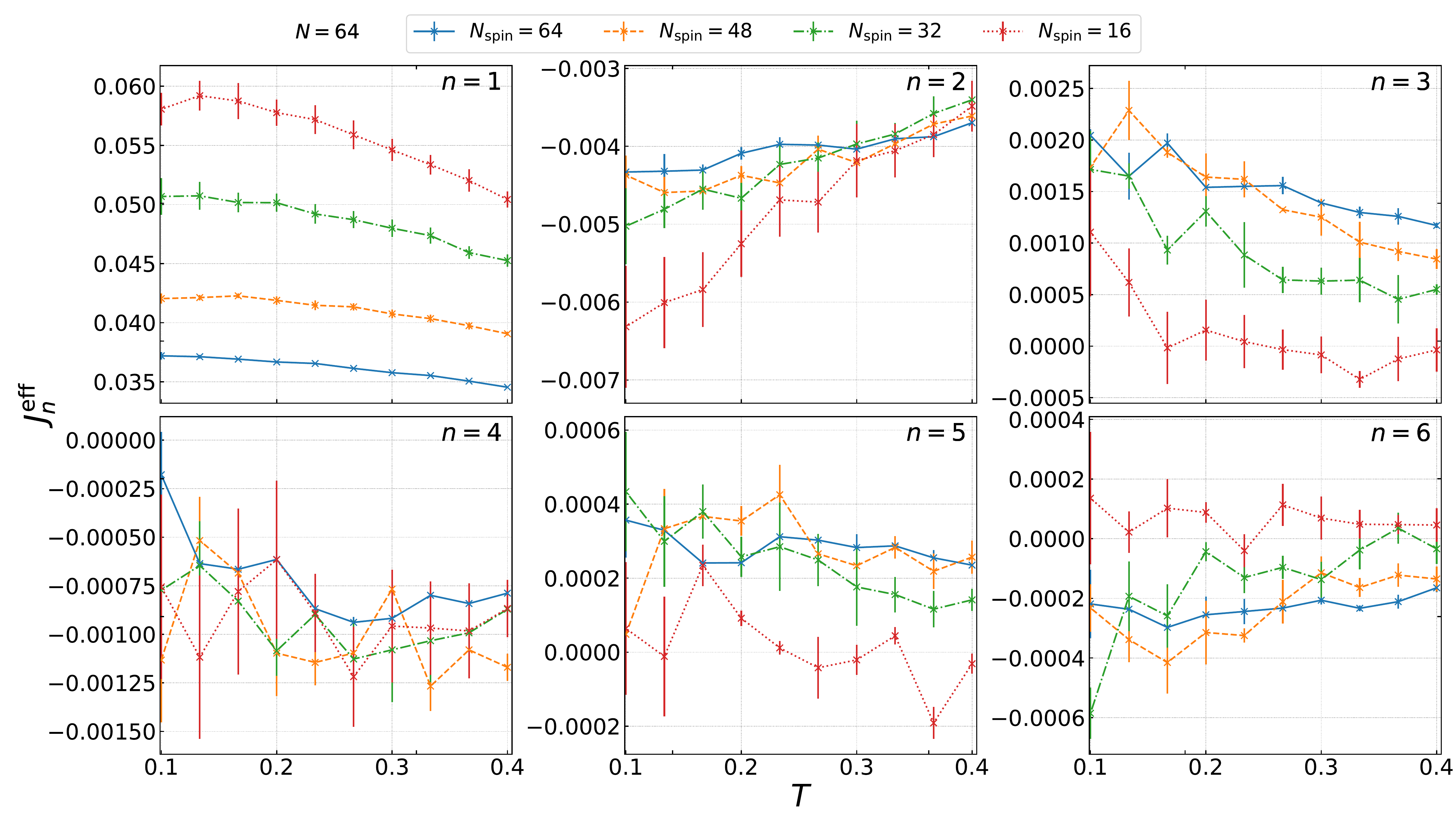}
 \caption{\label{fig:TJ}(Color online) The coupling constant $J_n^{\mathrm{eff}}$ of the trained effective models versus temperature for the regular and diluted models with various spin concentrations.
 The error bar is defined as the standard error over different localized spin configurations.
 In the case of the nearest-neighbor interaction $n=1$, the temperature dependence is relatively monotonic. The sample dependence appears to increase with decreasing spin concentration.}
\end{figure*}

In the regular DE model reported in the previous paper~\cite{SLMCDE}, 
it was found that the effective coupling constants oscillate and decay as a function of the distance between localized spins, which is similar to the RKKY interactions. 
In general, the RKKY interaction derived from the perturbation approach depends on temperature, and the amplitude of the oscillation is dumped at high temperature~\cite{RKKYT}
The RKKY interaction also depends on the localized spin configuration in general.
For example, on the bipartite lattice at half-filling, RKKY interaction between localized spins on the same or the opposite sublattice is always ferromagnetic or antiferromagnetic, respectively~\cite{RKKYP}. 
We investigate the spin concentration and temperature dependence of the effective model with the use of the SLMC, which the previous paper does not explore.

In Fig.~\ref{fig:rJ}, we show the coupling constant $J_n^{\mathrm{eff}}$ of the trained effective model against distances between localized spins at $T=0.4$ and $T=0.1$ with different spin concentrations . 
In the Table~\ref{table:1} ($T=0.4$) and the Table~\ref{table:2} ($T=0.1$), we summarize the obtained values of $J_n^{\mathrm{eff}}$. 
Here, $n=1$ means the nearest neighbor coupling and $n=2$ the next-nearest neighbor coupling, and so on.

As shown in Fig.~\ref{fig:rJ}, we find the RKKY features, oscillating and decaying behavior as a function of the distance between localized spins, for all spin concentrations at both a higher temperature $T=0.4$ and a lower temperature $T=0.1$. 
However, the amplitude of the coupling constant of the effective model depends on the spin concentration and the temperature.

We show the temperature dependence of the effective coupling constants $J_n^{\mathrm{eff}}$ in the trained effective models for each distance $n$ with different spin concentrations, as shown in Fig.~\ref{fig:TJ}. 
The characteristics are different for the nearest neighbor $n=1$ and other cases. 
For the nearest neighbor $n=1$, temperature dependence is relatively smaller with small error bars. 
Monotonic increase with decreasing temperature is consistent with the previous study~\cite{RKKYT}
On the other hand, for other cases, temperature dependence is nonmonotonic and relatively larger.
The number of further interactions is fewer than nearer ones. Therefore, they are relatively irrelevant and can easily fluctuate. 
Since the temperature dependence of the effective coupling constants is not so strong, 
the "annealing" procedure used to obtain the effective model in lower temperature works well in the site-diluted DE model.

As seen in the acceptance rates calculated, the temperature dependence of the effective model is small.
On the other hand, we observed temperature-dependent behavior in the effective coupling constant trained.
The effective model $\mathcal{H}_\mathrm{eff}$ is optimized minimizing MSE$_\mathcal{H}$ and satisfying $\mathcal{H}_\mathrm{eff}(\mathcal{S},\beta) \simeq \frac{\log W(\mathcal{S},\beta)}{\beta}$. 
This means an effective model should mimic not energy but local ``free energy'' at finite temperature.
The ``entropy'', therefore, may cause temperature-dependent behavior of effective models. 

There is another possible explanation for the existence of temperature dependence. 
In terms of the quantum field theory, one has to integrate out the electron degrees of freedom to obtain effective classical interactions. 
Two classical spins interact with each other through electron Green's functions in the effective Lagrangian. 
At zero temperature, such a Green's function has been obtained as a result of the exact summation of the Born series~\cite{Rusin}. 
At finite temperature, the electron Green's function depends on temperature. 
The SLMC imitates this effective model in the Hamiltonian formalism with the use of the temperature-dependent coupling constant. 

We show that the SLMC works well in random systems, and there is a simple effective model.
There are two ways to study the SG transition in the site-diluted DE model. 
One is the MC simulation with effective two-body interactions obtained by the SLMC. 
We find that, even in the strong coupling regime, there is an effective two-body interaction to describe the original DE model. 
Therefore, without doing the Metropolis-Hastings test, one might have good accuracy in the classical MC simulation with this effective interaction. 
The other is the SLMC with the use of the two-body classical spin interactions. 

To grasp the tail of the SG transition, one has to calculate systems in various sizes with fixed spin concentration, which takes much time. 
By accelerating the MCMC simulations with the use of the SLMC, one can study whether the SG transition occurs in the DE model or not, which is still an open question. 
We note that the transition temperature of the SG transition would be much lower than that of the ferromagnetic transition due to spin frustrations.
For the challenge to the problems of the SG transition in the itinerant electron systems, larger system sizes and much lower-temperature simulations are needed.
The temperature exchange method~\cite{HukushimaNemoto} is usually used for the classical localized SG models and will make our simulation better, but combinations of the temperature exchange method and the SLMC method are future work.

\section{Summary}
We performed the self-learning Monte Carlo methods to the semiclassical site-diluted double-exchange model.
With decreasing localized spin concentrations, we observed modest acceptance rates and, the effective models kept RKKY behavior, which oscillates and decays with distance.
As well as the regular DE model, the SLMC method works well with the diluted DE model.
We found that effective RKKY-like interaction depends on the spin concentrations and temperature.
We showed that the SLMC works well in random systems and, there is a simple effective model in a strong coupling regime.

\section*{Acknowledgment}
The calculations were partially performed by the supercomputing systems SGI ICE X at the Japan Atomic Energy Agency.
This work was partially supported by
JSPS--KAKENHI Grant Numbers 18K03552 to Y.N., and the ``Topological Materials Science'' (No. 18H04228) JSPS--KAKENHI on Innovative Areas to Y.N..

\appendix
\section{Details of Optimization Procedure of Effective Models}
\label{sec:opt}
In this appendix, we will explain datails how to optimize the effective models. We assume the effective model as $\mathcal{H}_{\mathrm{eff}}(\mathcal{S},\beta)=\sum_{\langle i, j\rangle_n} J^\mathrm{eff}_n(\beta) \bm{S}_i\cdot\bm{S}_j(\mathcal{S})$ and consider up to $N_{J^\mathrm{eff}}$ th neighbor interactions at inverse temperature $\beta$.

We prepare $M$ states $\mathcal{S}_1, \dots, \mathcal{S}_M$ as training data and $C_n(\mathcal{S})\equiv\sum_{\langle i, j\rangle_n} \bm{S}_i\cdot\bm{S}_j(\mathcal{S})$. A matrix $C$ and a vector $\bm{J}$ are defined as follows:
\begin{align}
 C &\equiv \left(
 \begin{array}{cccc}
 1 & C_1(\mathcal{S}_1) &\cdots& C_{N_{J^\mathrm{eff}}}(\mathcal{S}_1) \\
 \vdots & \vdots &\ddots& \vdots \\
 1 & C_1(\mathcal{S}_M) &\cdots& C_{N_{J^\mathrm{eff}}}(\mathcal{S}_M)
 \end{array}
 \right),\\
 \bm{J}&\equiv\left(
 \begin{array}{c}
 E_0 \\
 J^\mathrm{eff}_1 \\
 \vdots \\
 J^\mathrm{eff}_{N_{J^\mathrm{eff}}}
 \end{array}
 \right).
\end{align}
We also define a vector $\bm{\mathcal{H}}\equiv{}^t\left(\mathcal{H}(\mathcal{S}_1), \cdots, \mathcal{H}(\mathcal{S}_M)\right)$ is aligned energies calculated via states as traning data. For the acutual simulation, you should restore $\bm{C}$ used in the following procedure.

$\bm{J}$ which give a minimum of $\|\bm{\mathcal{H}}-C\bm{J}\|^2$ is a solution of the normal equation ${}^tCC\bm{J}={}^tC\bm{\mathcal{H}}$ and it becomes $\bm{J}=\left({}^tCC\right)^{-1}{}^tC\bm{\mathcal{H}}$. The number of learning parameters to train is a few (only $N_{J^\mathrm{eff}}$ parameters), a direct calculation of the $N_{J^\mathrm{eff}}+1$ dimensitonal inverse matrix $\left({}^tCC\right)^{-1}$ is not expensive. 
Of course, if the number of learning parameters is large, \textit{e.g.}, an effective model using the deep neural network~\cite{SLMCDeep}, you should use an iterative method.

\section{Temperature Dependence of Filling for $J=16$}\label{sec:fill}
\begin{figure}
    \includegraphics[width=\columnwidth]{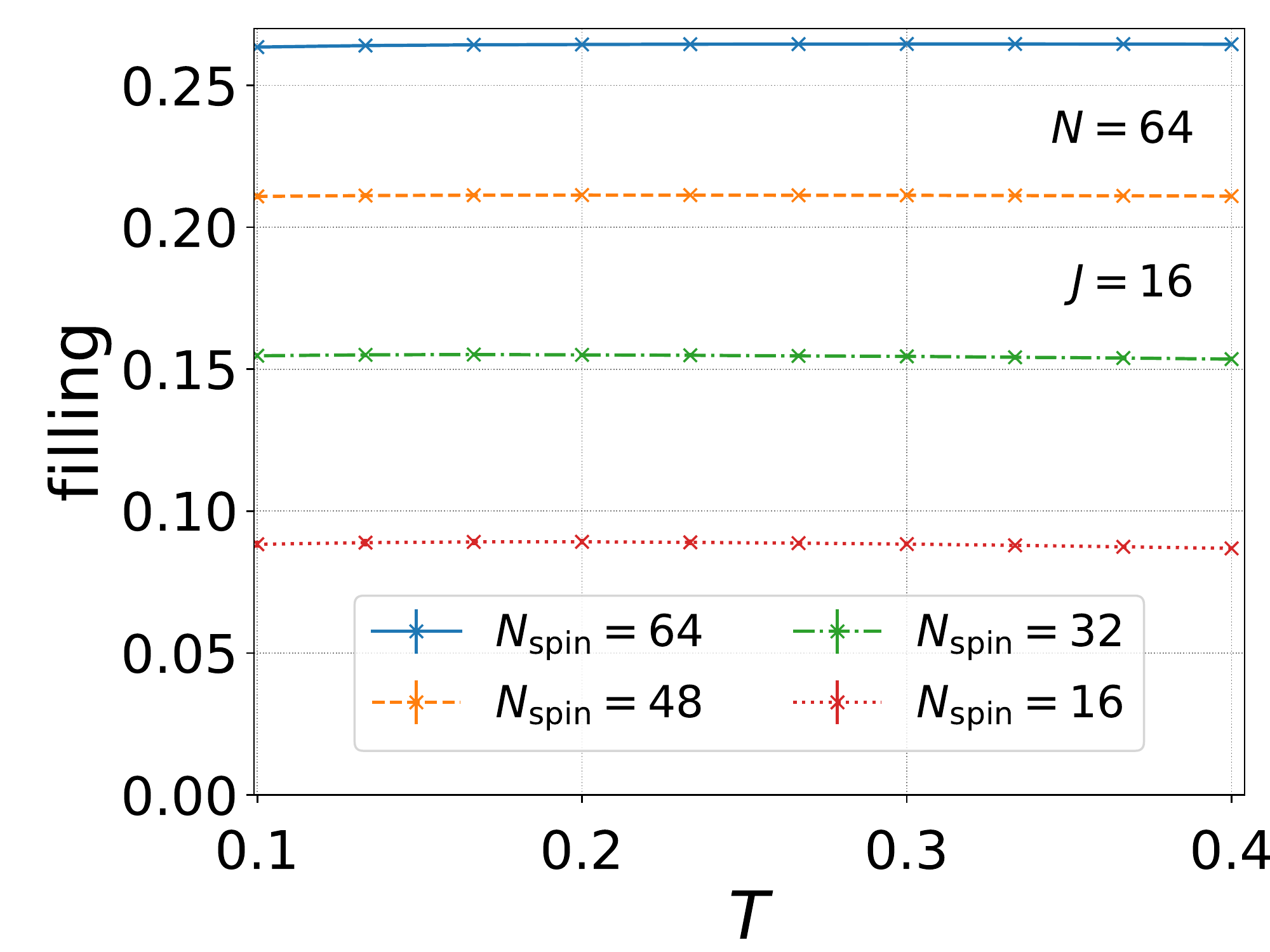}
    \caption{\label{fig:TNJ16}(Color online) The filling versus temperature with different spin concentrations.
    The error bar is defined as the standard error over different localized spin configurations.
    The size of the errorbars is smaller than that of markers. 
    The filling has little temperature dependence and is proportional to the spin concentration.}
\end{figure}
In this appendix, we discuss the filling for $J=16$ in the regular and diluted DE model.
We show the result of the filling in Fig.~\ref{fig:TNJ16}.
Numerical conditions are the same as the ones in the main text.
The values of filling decrease monotonically with decreasing the spin concentration $N_\mathrm{spin}$ and are almost independent of temperature.
The values of filling seem to be proportional to $N_\mathrm{spin}$.
This is naturally explained by the assumption that one localized spin gathers the same amount of electrons, and the physics around localized spin is similar to each other.

\section{$J$ Dependence of Effective Models}
\begin{figure}
    \includegraphics[width=\columnwidth]{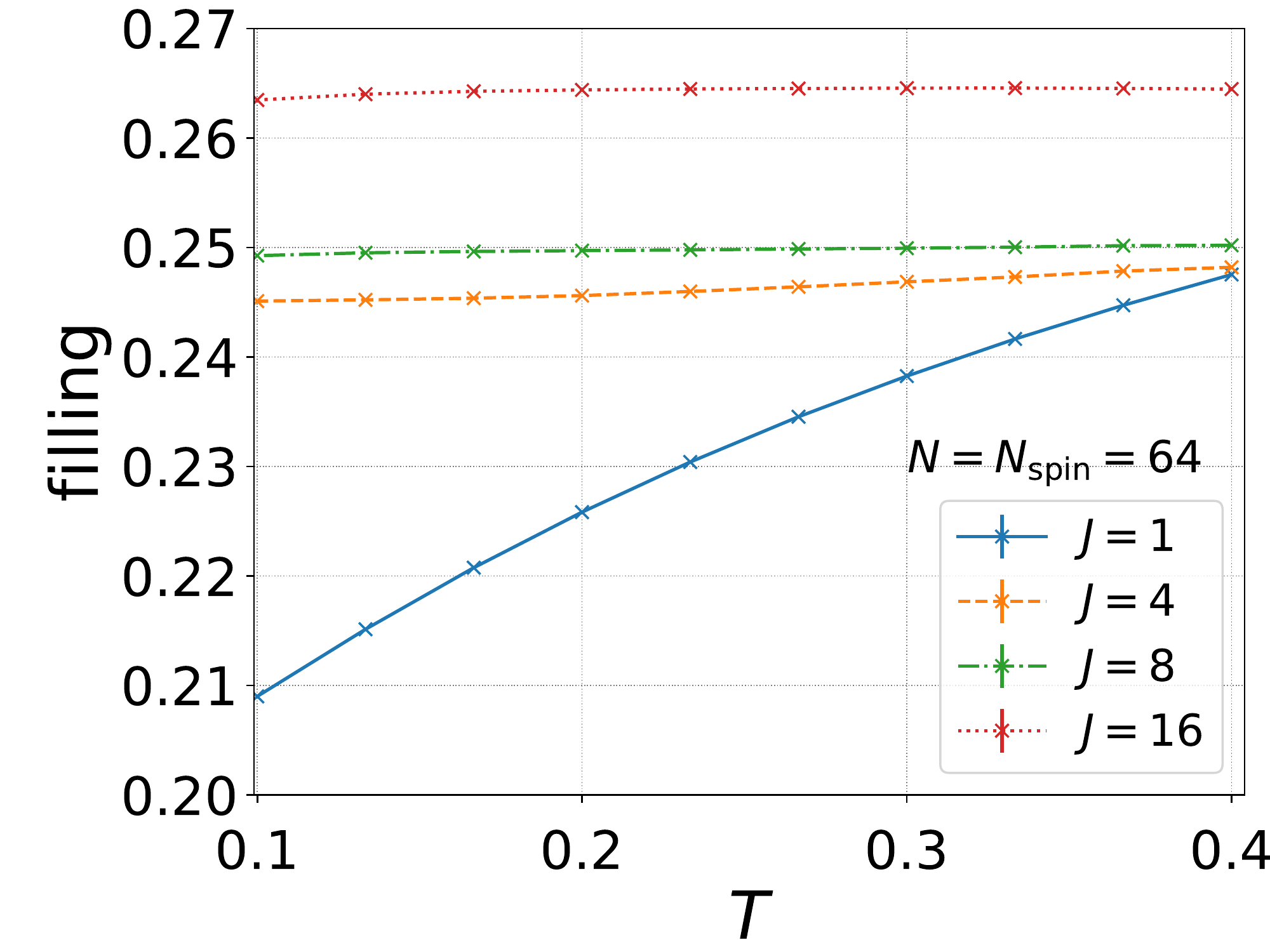}
    \caption{\label{fig:TN}(Color online) The filling versus temperature with different $J$ in the regular DE model.
    The error bar is defined as the standard error over different localized spin configurations.
    The size of the error bars is smaller than that of markers. 
    The chemical potential $\mu$ differs among $J$ and is chosen so that the filling is close to the quarter filling.
    Only for the case of $J=1$, the filling depends on temperature.}
\end{figure}
\begin{figure}
    \includegraphics[width=\columnwidth]{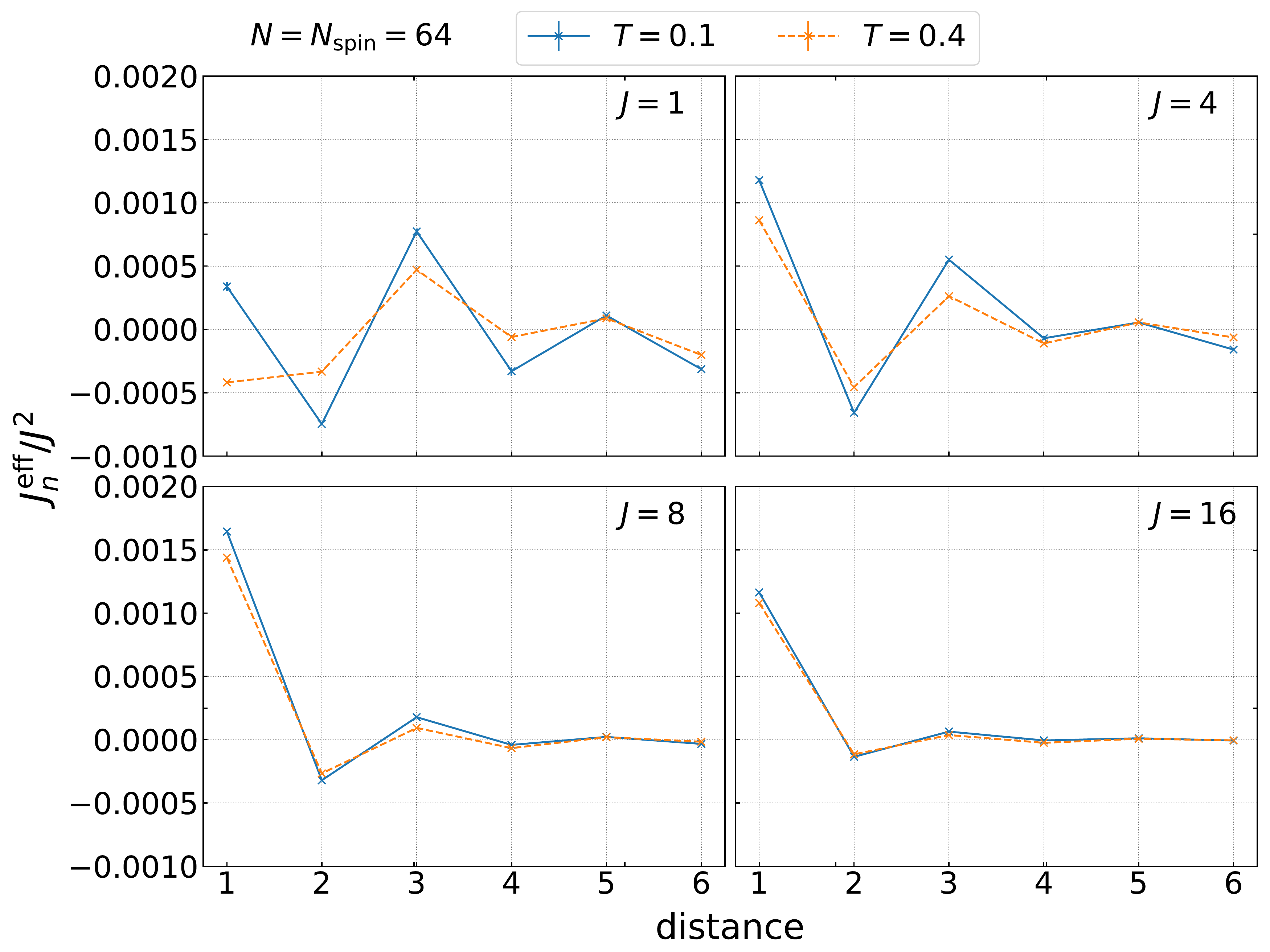}
    \caption{\label{fig:rJoJ2}(Color online) The coupling constant $J_n^{\mathrm{eff}}/J^2$ of the trained effective model versus distance between localized spins with different $J$ at $T=0.1$ and $T=0.4$ in the regular DE model.}
\end{figure}
\begin{figure*}
    \includegraphics[width=2\columnwidth]{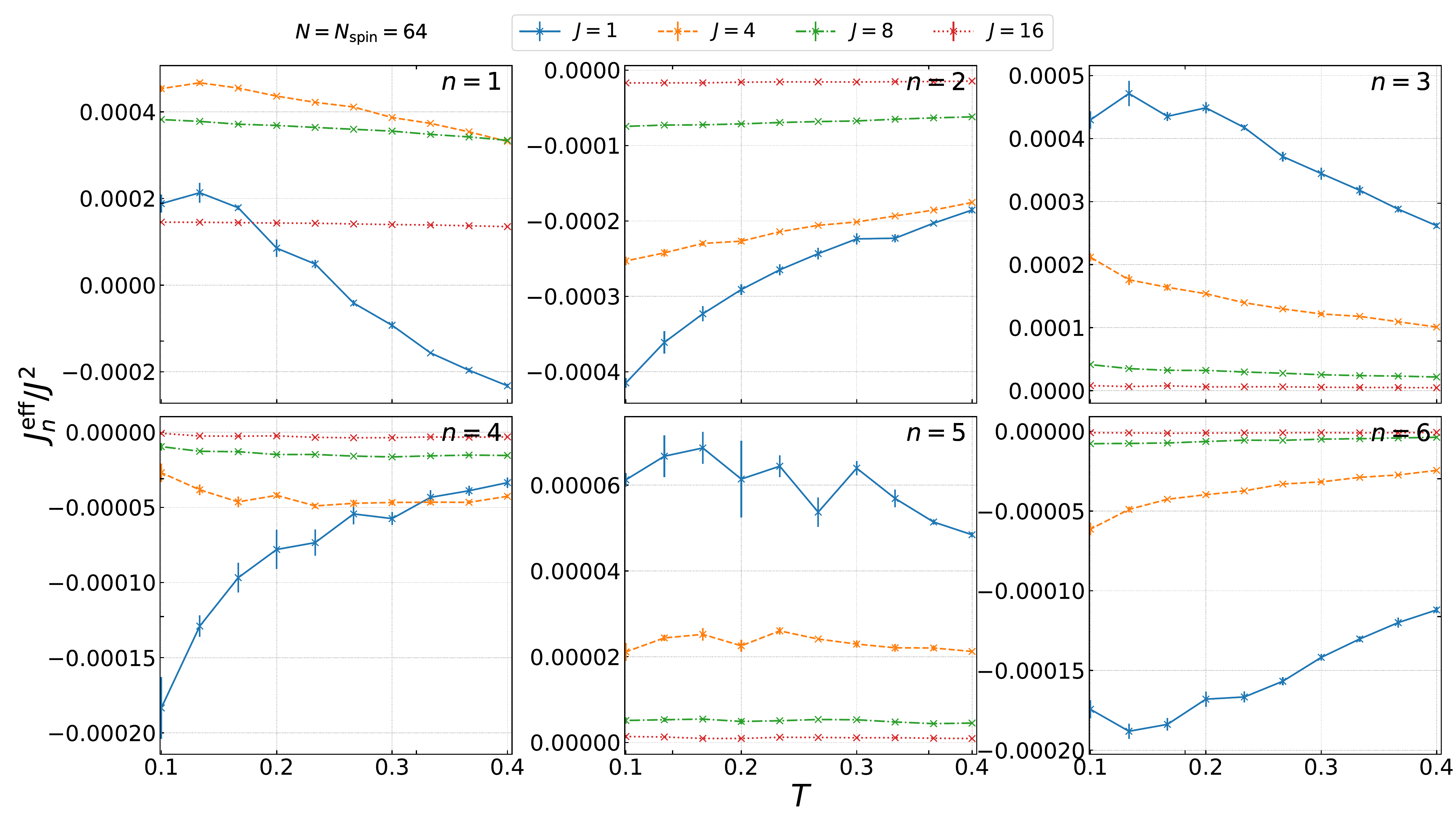}
    \caption{\label{fig:TJroJ2}(Color online) The coupling constant $J_n^{\mathrm{eff}}/J^2$ of the trained effective models versus temperature in the regular DE model. The error bar is defined as the standard error over different localized spin configurations.}
\end{figure*}
In the main text, we set $J=16$ because a sufficiently large $J$ is needed for the ferromagnetic transition to appear in the regular DE model.
On the other hand, the RKKY interaction is derived by the second-order perturbation with respect to $J$, which means the model with RKKY interaction is better to approximate the DE model when $J$ is sufficiently small.

To investigate the relationship between the magnitude of $J$ and the effective spin-spin interaction $J_n^\mathrm{eff}$,
we trained the effective models with $J=1, 4, 8$ in the regular DE model.
First, in Fig.~\ref{fig:TN}, We show the result of the filling for $J=1, 4, 8$, and $16$.
The chemical potential $\mu$ is chosen to be close to the quarter filling at $T=0.4$; $\mu=-1.8, -2.6, -4.3$, and $-8$ for $J=1, 4, 8$, and $16$, respectively.
Except for $J=1$, the fillings are almost independent of temperature.
On the other hand, for $J=1$, the filling start from about $0.25$ at $T=0.4$ to about $0.21$ at $T=0.1$.

Next, we show the results of $J_n^\mathrm{eff}$ as the function of the distance between localized spins in Fig.~\ref{fig:rJoJ2}.
We find that the amplitude of $J_n^\mathrm{eff}/J^2$ normalized by $J^2$ is similar in different $J$.
The oscillation with distance is also observed in all $J$.
Since RKKY interaction derived from a second-order perturbation theory has a factor $J^2$ and algebraic decay with oscillation, 
this suggests that the RKKY-like description is not bad in a wide range of interaction strengths.
With the use of the SLMC, we find that the effective model that has been initially derived from a perturbation theory can be used beyond its physical validity, which has also been observed in an application of the SLMC method to the lattice quantum chromodynamics calculation~\cite{SLMCQCD}.

We show the temperature dependence of the effective coupling constants $J_n^{\mathrm{eff}}$ in the trained effective models for each distance $n$ in Fig.~\ref{fig:TJroJ2}. 
The temperature dependence of coupling constants is weak except for the $J=1$ case. 
One of the differences between $J=1$ case and others is the temperature dependence of the filling as shown in Fig.~\ref{fig:TN}. 
Generally the RKKY interaction depends on the Fermi wave vector, which is changed by the electron filling. 
This suggests that the temperature dependence of the $J_n^{\mathrm{eff}}$ for $J=1$ originates from the temperature dependence of the filling.

\section{Effective Model Near Transition Temperature}\label{sec:tc}
\begin{figure*}
    \includegraphics[width=2\columnwidth]{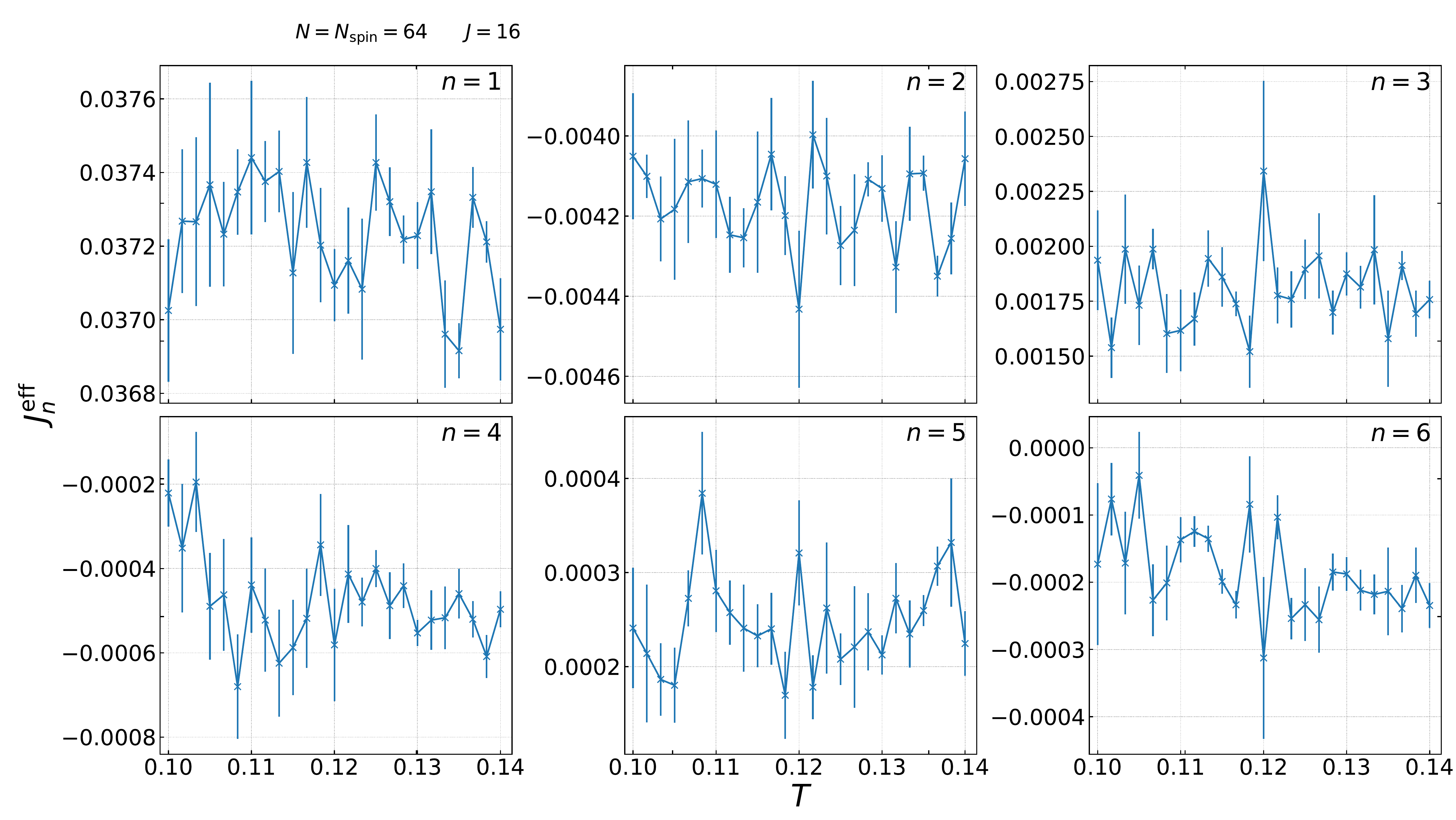}
    \caption{\label{fig:TJTc}(Color online) The coupling constant $J_n^{\mathrm{eff}}$ of the trained effective models versus temperature in the regular DE model.
    The error bar is defined as the standard error over different localized spin configurations.
    $J_n^{\mathrm{eff}}$ is continuously connected below and above $T_\mathrm{c}\simeq0.12$}
\end{figure*}

In the regular DE model, the ferromagnetic transition appears at $T_\mathrm{c}\simeq0.12$~\cite{MotomeFurukawa}.
We check the temperature dependence of the effective model near the critical temperature $T_\mathrm{c}$ in detail.
We use the same parameters used in the main text. 
We change the learning process in this appendix. The effective model is learned from initial random numbers, and the temperature is fixed in the same simulation. 
In this learning process, the effective models with different temperatures are independently obtained. 
Fig.~\ref{fig:TJTc} shows that no singular behavior occurs around the critical temperature $T_\mathrm{c}$. 

\clearpage
\bibliography{69739}
\clearpage
\end{document}